# Contribution à la modélisation et simulation numérique de l'écoulement du sang dans l'artère

H. ALLA[a], M.S. GUEBLAOUI[a], M.H. BENSAID[a]

*a-Laboratoire de modélisation et simulation. Faculté des sciences, BP 1505 El M'Naour Bir el Djir Oran, 31000, Algérie.*

**Résumé :**

*Nombreux sont les interrogations soulevées par la médecine interventionnelle, concernant la prise en charge des pathologies de la paroi artérielle (anévrisme, dissection, coarctation, athérosclérose) .pour cela nous avons fait la modélisation et la simulation numérique de l'écoulement du sang dans l'artère rénale prise par l'Imagerie Médical. La géométrie a été reconstruite à partir des images médicales d'angiographie, l'angioscanner, l'écho-Doppler et l'IRM. En ne considérant que le sang comme un fluide newtonien et écoulement stationnaire. Les résultats obtenus en termes des paramètres physiques tels que la vitesse et la pression dynamique sont montrés que le cas le plus simple suffit pour accéder à des paramètres susceptibles d'expliquer les phénomènes de sténose ou de thrombose dans les artères.*

**Abstract:**

*Numerous are the questionings raised by medicine interventionnelle, concerning the hold in charge of the pathologies of the arterial partition (aneurysm, dissection, coarctation, atherosclerosis).for it we made the modeling and the numeric simulation of the blood flow in the renal artery taken by the Medical imagery. Geometry has been rebuilt from the medical pictures of angiography, angioscanner and IRM. While considering that blood like a fluid Newtonian and stationary flow. The results gotten in terms of the physical parameters as the velocity, the dynamic pressure is shown that the simplest case was enough to collect relevant data for the development of stenos or thrombosis in the arteries.*

**Mots clefs : Biomécanique des fluides, artère rénale, paroi artérielle, sténose rénale**

## 1 Introduction

De nombreuses études ont été effectuées sur l'écoulement sanguin dans les systèmes artériel, avec des données physiques et pathologique, dans le but de trouver des réponses aux maladies et complications. Pour mener une étude numérique permettant de comprendre et définir les facteurs essentiels influençant sur l'écoulement sanguin dans une artère et afin d'étudier la relation entre l'écoulement du sang et la physiopathologie, il est nécessaire de connaître ses caractéristiques locales dans les artères.

L'artère rénale est une artère issue de l'aorte abdominale et irrigant le rein. Il existe deux artères rénales chez l'homme, l'artère rénale droite et gauche. Chacune apporte du sang oxygéné au rein homolatéral. Elles naissent de l'aorte au niveau de la vertèbre T12 et se dirigent vers le bas et le dehors. Elles émettent une artère surrénale à destination des glandes surrénales puis se divisent en plusieurs artères, artères surrénales vers le haut et en artères urotéliques vers le bas, avant d'entrer par le hile du rein. FIG. 1

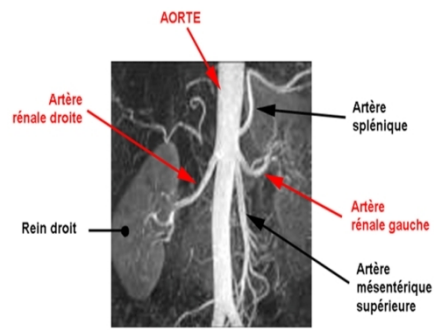

FIG. 1 – Artériographie des artères rénale.

La pathologie vasculaire du rein est une pathologie variée dont le diagnostic a été longtemps l'apanage quasi-exclusif de l'artériographie. Actuellement, les avancées techniques ont permis un réel essor des moyens d'imagerie en coupes : l'échographie doppler, l'angioscanner et l'angio-IRM qui sont devenues de plus en plus disponibles. Ces techniques permettent de faire le diagnostic des pathologies rénovasculaires avec de bonnes spécificités et sensibilités. Tous ces progrès techniques ont restreint la place de l'artériographie diagnostique. Parallèlement, l'évolution des techniques de cathétérisme a été à l'origine du développement de la radiologie interventionnelle endovasculaire.

La sténose artérielle est une maladie fréquente, qui se rapporte à un rétrécissement d'une artère, dû au dépôt de plaque d'athéroscléroses sur les parois interne du vaisseau, et qui peut entraver partiellement ou totalement l'écoulement dans le réseau artériel.

La sténose de l'artère rénale est définie par le rétrécissement du calibre de l'artère rénale et peut être uni- ou bilatéral. Cet état est souvent asymptomatique mais peut provoquer une hypertension dite reno-vasculaire ou une insuffisance rénale sur ischémie. L'imputation d'une hypertension artérielle à une sténose d'artère rénale ne peut en fait être démontrée formellement que si la correction de l'obstacle normalise la pression artérielle. La sténose artérielle rénale n'est à l'origine que de 1 à 3% des cas d'hypertension. FIG. 2

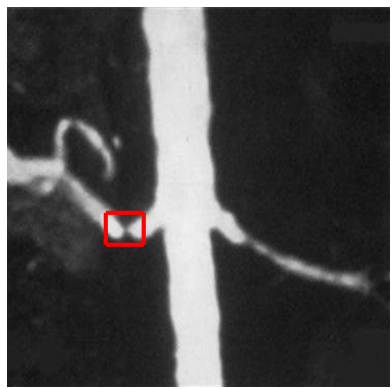

FIG. 2 – Sténoses d'une artère rénale, en rouge on voit bien la sténose.

## 2  Méthodes

## 2.1  Modèle géométrique

Pour accéder à l'anatomie réelle, le seul moyen non invasif repose sur l'accès aux imageries médicales (l'Angioscanner). L'objectif est de construire une géométrie en 3D qui corresponde à la forme réelle de l'artère rénale à l'aide de logiciel Gambit 1.3.0, nous avons ajouté un cylindre droit à l'entrée de l'artère suffisamment long pour bien établir l'écoulement et l'autre à la sortie pour éviter la perturbation de l'écoulement. FIG. 3

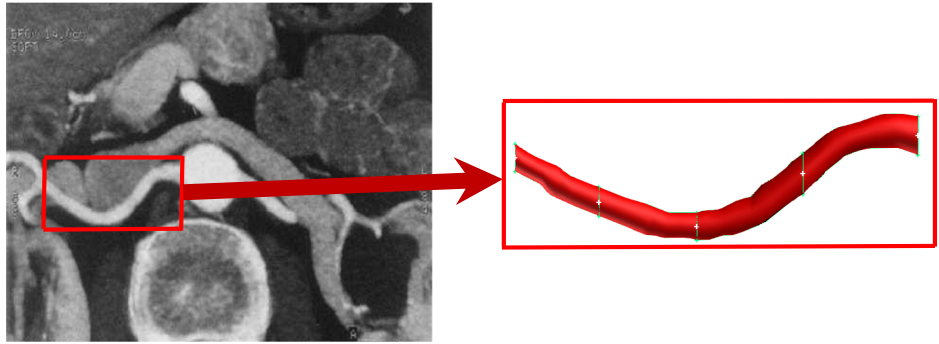

FIG. 3 – Construction en 3D de l'artère rénale droit.

## 2.2  Maillage

La géométrie est maillée par hexaédrique ce maillage est extrudé à l'aide de l'outil Cooper proposé par Gambit 1.3.0, cette méthode nous permet d'obtenir des mailles qui sont orientées suivant la morphologie de l'artère et donc dans le sens de la vitesse.

La meilleure méthode pour vérifier le maillage est la réalisation de la simulation numérique que l'on pourra effectuer sur le logiciel Fluent 6.2.16. (FIG. 4) et TAB. 1.

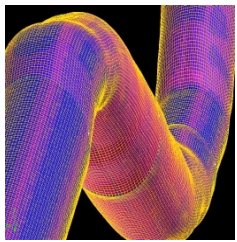

FIG. 4 – Maillage de la géométrie.

| Nombre | Artère normale | | Artère touché | |
|---|---|---|---|---|
| Cellules | 335400 | | 563418 | |
| faces | 1029719 | | 1720633 | |
| Nœuds | 359520 | | 594405 | |
| Volume total ($m^3$) | 3.727147e+04 | | 4.318282e+04 | |
| | Max | Min | Max | Min |
| X-Coordonnée (m) | 1.000000e+00 | 2.500000e+02 | 1.000000e+00 | 2.500000e+02 |
| Y-Coordonnée (m) | 1.245796e+02 | 1.790448e+02 | 1.239620e+02 | 1.791618e+02 |
| Z-Coordonnée (m) | -9.493029e+00 | 9.493029e+00 | -8.996049e+00 | 8.996049e+00 |
| Volume ($m^3$) | 7.935349e-03 | 2.632265e-01 | 1.461309e-02 | 1.698167e-01 |
| Face surface ($m^2$) | 1.995973e-02 | 7.423854e-01 | 3.881219e-02 | 3.807629e-01 |

TAB. 1 – Caractéristique de maillage

## 2.3 Simulation numérique

L'étude des contraintes mécaniques et hémodynamiques permette mieux de comprendre les pathologies des vaisseaux tels que la sténose et thrombose, et proposer des améliorations thérapeutiques.

L'écoulement sanguin dans l'artère est calculé à partir des équations de Navier-stokes résolues à l'aide de Fluent 6.2.16 par la méthode des volumes fins. Dans ce type des artères, le sang est considéré comme un fluide newtonien de viscosité 0,004, de masse volumique 1050 kg /$cm^3$ et de vitesse d'entré 0.3m/s. Les parois de l'artère rénale sont traitées dans la simulation comme solides.

La simulation permet de visualiser les lignes de courant colorées en fonction de champ de vitesse,
le profile de vitesse et de la pression dynamique. FIG. 5

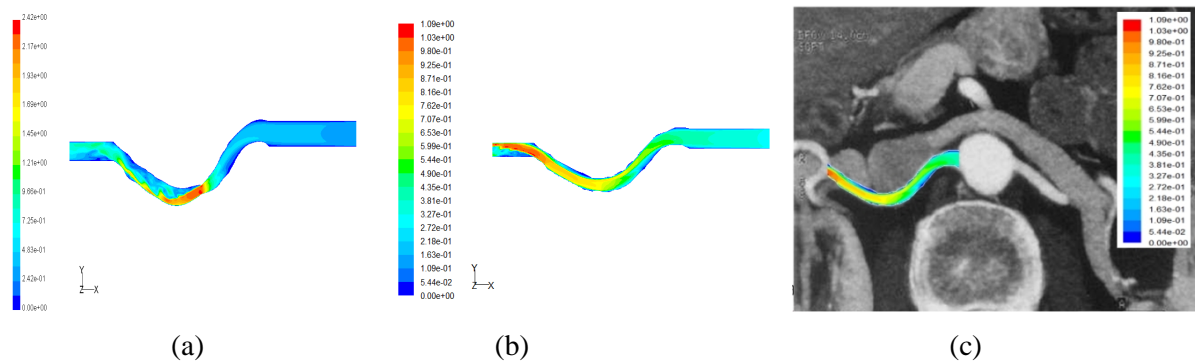

(a)          (b)          (c)

FIG. 5 – Contour de vitesses dans les deux cas de l'artère sténosé (a) normale(b).

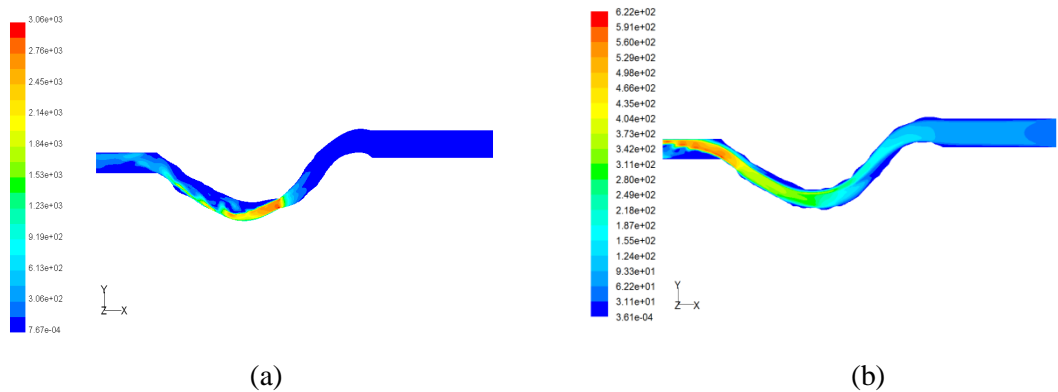

(a)                                                           (b)

FIG. 6 – Contour de pressions dynamique dans les deux cas de l'artère sténosé (a) normale(b).

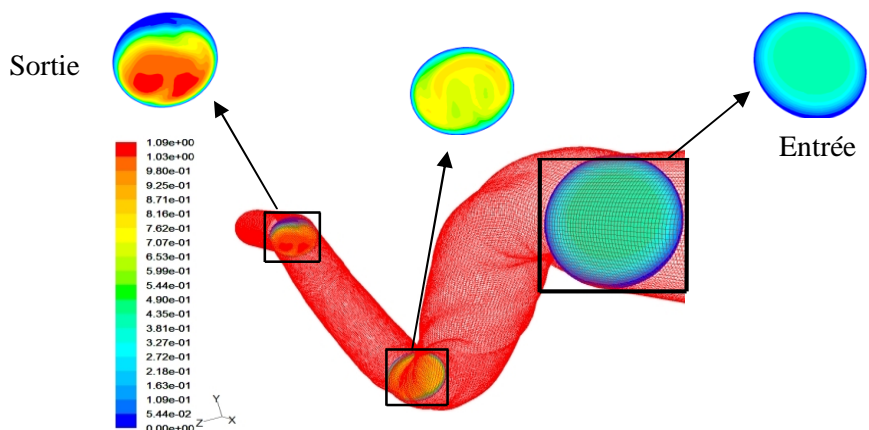

FIG. 7 – représente contour de vitesse de l'artère sténosé.

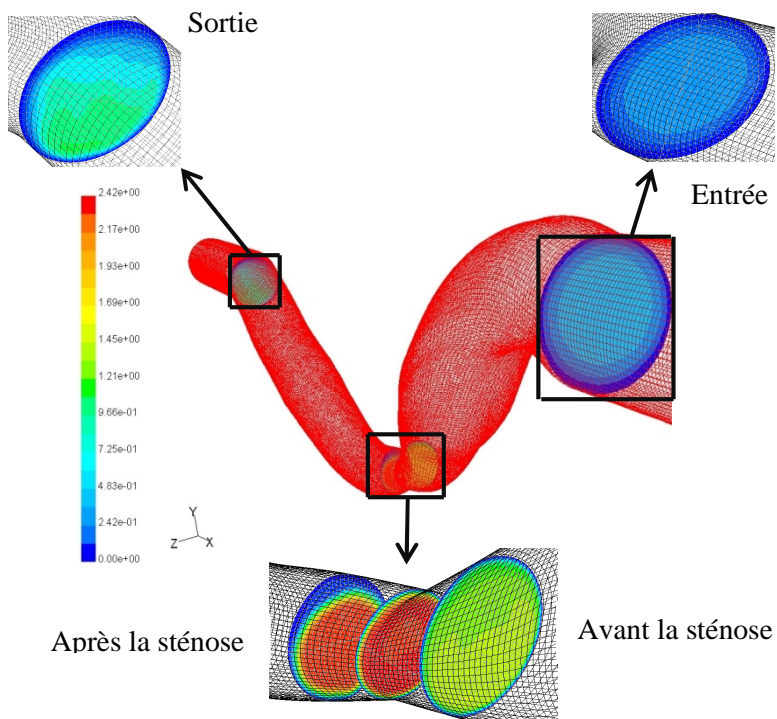

FIG. 8 – Contour de vitesses de l'artère sténosé à des régions spécifiées.

## Conclusion

L'étude numérique menée nous montre bien la concordance des résultats avec la réalité. En effet le contour de vitesses après et avant la sténose illustré en FIG. 7 et FIG. 8 traduit bien le cas réel, de même que le contour de pression FIG. 6.Ainsi donc il est plus que nécessaire pour la chirurgie une telle étude, dans le but de trouver des réponses aux maladies et complications liées aux artères.